\documentclass[12pt,letterpaper]{JHEP3}         
\usepackage{amssymb,amsfonts}

\usepackage{epsfig}

\def\be{\begin{eqnarray}}
\def\ee{\end{eqnarray}}

\newcommand\para{\paragraph{}}

\newcommand{\eqn}[1]{(\ref{#1})}

\def\Dslash{\,\,{\raise.15ex\hbox{/}\mkern-12mu D}}
\def\Dbarslash{\,\,{\raise.15ex\hbox{/}\mkern-12mu {\bar D}}}
\def\delslash{\,\,{\raise.15ex\hbox{/}\mkern-9mu \partial}}
\def\delbarslash{\,\,{\raise.15ex\hbox{/}\mkern-9mu {\bar\partial}}}
\def\pslash{\,\,{\raise.15ex\hbox{/}\mkern-9mu p}}
\def\calDslash{\,\,{\raise.15ex\hbox{/}\mkern-12mu {\cal D}}}

\newcommand{\ZZ}{{\mathbb Z}}

\newcommand{\RR}{{\mathbb R}}

\newcommand{\ads}{AdS${}_4$\ }

\def\lae{\mathrel{\mathop{\smash{\lower .5 ex \hbox{$\stackrel<\sim$}}}}}
\def\lae{\mathrel{\mathop{\smash{\lower .5 ex \hbox{$\stackrel>\sim$}}}}}

\title{
\vspace{-22mm}
\hfill{\small \tt WIS/14/12-AUG-DPPA}\vskip 50pt
{\ }\\{\ }\\
Confinement in Anti-de Sitter Space}

\author{Ofer Aharony${}^{1,2}$, Micha Berkooz${}^1$,
David Tong${}^{3,4}$ and Shimon Yankielowicz${}^5$ \\

${}^1$ Department of Particle Physics and Astrophysics,
Weizmann Institute of Science, Rehovot 76100, Israel \\
${}^2$ School of Natural Sciences, Institute for Advanced Study, Princeton, NJ 08540, USA\\
${}^3$ Department of Applied Mathematics and Theoretical Physics, University of Cambridge, UK\\
${}^4$ Albert Einstein Minerva Center, Weizmann Institute of Science, Rehovot 76100, Israel \\
${}^5$ School of Physics and Astronomy,
The Raymond and Beverly Sackler Faculty of Exact Sciences,
Tel Aviv University, Ramat Aviv, 69978, Israel\\ {\ } \\
{\tt E-mails : Ofer.Aharony@weizmann.ac.il,  Micha.Berkooz@weizmann.ac.il, D.Tong@damtp.cam.ac.uk,
 Shimonya@post.tau.ac.il}
}

\abstract{Four dimensional gauge theories in anti-de Sitter space, including pure Yang-Mills theory,
exhibit a quantum phase transition between a deconfined phase and a confined phase as the gauge coupling is varied. We explore various mechanisms by which this may occur, both in a fixed background and in the presence of gravity. We also make a number of observations on the dynamics of four dimensional supersymmetric gauge theories in anti-de Sitter space.}

\begin{document}
\pagestyle{plain} \setcounter{page}{1}
\newcounter{bean}
\baselineskip16pt

\section{Introduction}

A quantitative understanding of the confinement mechanism  in four-dimensional non-Abelian gauge theories remains elusive \cite{Greensite:2011zz}. Part of the difficulty is that, away from the large $N$ limit, there are no small parameters that one can use to attack the problem.

\para
In this paper, we approach the problem by considering the dynamics of Yang-Mills fields in a curved background spacetime given by the four dimensional anti-de Sitter space AdS${}_4$. As will become clear, this doesn't obviously make the problem any easier. However, it does arguably make the problem somewhat cleaner. The main advantage was pointed out long ago by Callan and Wilczek: anti-de Sitter  tames the infra-red \cite{cw}. At distances large compared to the AdS radius $L$, the classical theory already has a mass gap, cutting off any infra-red (IR) divergences. Furthermore, the running of the gauge coupling $g^2(\mu)$ halts at the energy scale $\mu \sim 1/L$. Unlike  a finite box, AdS space cuts off IR divergences without reducing the symmetries; the dimension of the $SO(3,2)$ isometry group of AdS${}_4$ is the same as that of the Poincar\'e group\footnote{An alternative way to control the infrared divergences and learn about confinement, which breaks some of the symmetries, is to compactify the theory on a circle, leading to an Abelian theory at low energies. This method \cite{Polyakov:1976fu} recently found several interesting new applications; see, for instance, \cite{Unsal:2007vu,Unsal:2007jx,Shifman:2008ja,Unsal:2008ch,Poppitz:2009tw,Poppitz:2011wy}.}.

\para
The upshot of this is that, in marked contrast to flat space, asymptotically free gauge theories in anti-de Sitter space are characterized by a dimensionless parameter which may be chosen to be $g^2(1/L)$ or, equivalently, $\Lambda L$ where $\Lambda$ is the strong coupling scale.
%
%
Our interest lies in theories which, like pure Yang-Mills, would confine when placed in flat space. We will attempt to follow how the physics changes as we  vary the dimensionless parameter  $\Lambda L$.  There are two extreme limits:
\begin{itemize}
\item {\bf Weak Coupling}, $\Lambda L \ll 1$: In this regime, the theory is weakly coupled at the AdS scale and the physics at all scales is controlled by the classical Yang-Mills action, with calculable perturbative corrections. When the theory is placed in global AdS$_4$, it has a mass gap of order $1/L$. The states are weakly coupled gluons.
\item {\bf Strong Coupling}, $\Lambda L \gg 1$: In this regime, as we go down in energy the strong coupling effects become important  before  the curvature of spacetime plays any role. The result is a confining gauge theory in AdS$_4$. Dynamical states are singlets of the gauge group, with a mass gap of order $\Lambda$.
\end{itemize}
As we vary $\Lambda L$, the theory undergoes a (de)confinement transition. Such a phase transition is familiar at finite temperature in flat space; here it arises as a zero temperature, quantum phase transition. Notice that this is qualitatively different from  compactifications of Yang-Mills theories on manifolds of finite volume where, at least at finite $N$,  phase transitions can never occur. The purpose of this paper is to explore the mechanism for this transition in AdS.

\para
The gauge/gravity correspondence means that dynamics in AdS spacetimes gains added interest. When coupled to gravity, there exists a dual interpretation of the bulk dynamics in terms of a three dimensional conformal field theory (CFT) \cite{Maldacena:1997re,Gubser:1998bc,Witten:1998qj}.  Gauge theories in anti-de Sitter space occur naturally in string theoretic constructions, where the non-Abelian symmetry may arise from stacks of D-branes or from reduction on an internal manifold.  To the best of our knowledge,
 discussion of the AdS/CFT correspondence in the literature is restricted to the regime where the fields in AdS space are weakly coupled. However, there is no reason to believe that the confining phase is affected by the addition of a dynamical, weakly coupled gravitational sector, and it is natural to ask how confinement manifests itself in the dual CFT. In the usual AdS/CFT dictionary, the bulk non-Abelian gauge field is dual to a non-Abelian global current $J_i(x)$ in the dual 3d CFT\footnote{As we describe in the next section, this statement is  true only once certain boundary conditions have been specified.}. At weak coupling, the  two-point function of currents is schematically of the form\footnote{The overall normalization of the currents is fixed by requiring that the corresponding charges satisfy the correct non-Abelian algebra, or, equivalently, by fixing the normalization of the term involving the structure constants $f_{abc}$ in the operator product expansion of two currents.}
\be \langle J_i(x)J_j(0)\rangle \sim \frac{1}{g^2}\left(\delta_{ij}\partial^2-\partial_i\partial_j\right)\frac{1}{x^2},\label{jj}\ee
with $g^2$ evaluated at $\mu = 1/L$.  Roughly speaking, the coefficient $1/g^2$ counts the number of charged fields in the 3d conformal field theory. (This statement can be made precise in 2d CFTs but not, apparently, in higher dimensions). We will discuss various scenarios for the behaviour of this two-point function as we go to strong coupling.


\para
In this paper, we take care to  distinguish between the cases of  gauge theories in a fixed AdS background and gauge theories in the presence of dynamical gravity.  (Few of the precise mathematical statements change, although this is primarily because we have few such statements to begin with. However, some of the words that we drape around the statements change). In Section \ref{basicsec}, we discuss some basic issues of the confinement transition in the absence of gravity. We will describe a number of mechanisms by which confinement could plausibly occur. We also discuss the behaviour of our theories at finite temperature. In Section \ref{turnonsec}, we revisit these issues in the presence of gravity, and examine the interpretation from the perspective of the dual conformal field theory.  Finally, in Section \ref{susysec}, we turn to a number of aspects of supersymmetric gauge theories in AdS, where greater control over the dynamics is available. Unfortunately, however, supersymmetry in AdS is less powerful than supersymmetry in flat space, and does not seem sufficient to shed light on the questions of confinement. An appendix includes some comments on instantons in AdS space.

\para
In this paper we will focus on four dimensional theories, though similar phenomena exist also in lower dimensions.
It would be interesting to generalize our discussion to other space-time dimensions. Note that this generalization is not straightforward because the boundary conditions for gauge fields on anti-de Sitter space are very different in different dimensions \cite{mr}. It would also be interesting to understand how other strong coupling phenomena, such as strongly coupled fixed points or dualities, manifest themselves in gauge theories on AdS space.

%
%
%
%

\section{Scenarios for Confinement}\label{basicsec}

We start by outlining a number of possible mechanisms for the confinement phase transition. Throughout this section we will decouple gravity completely and talk only of gauge dynamics in a fixed AdS background. We work in global \ads in Lorentzian signature with boundary ${\bf S}^2\times {\bf R}$. We focus on the pure Yang-Mills theory with gauge group $SU(N)$, though most of our discussion applies to any asymptotically free gauge theory which confines in flat space.

\subsubsection*{Weak Coupling}

To define the theory in AdS, we must pay attention to the boundary conditions. When $\Lambda L \ll 1$, we are dealing with  weakly coupled $SU(N)$ gauge fields in AdS${}_4$. There are two possible boundary conditions that one can impose \cite{hr,wittensl2z,mr}: electric (Dirichlet) and magnetic (Neumann)\footnote{Additional choices are also possible \cite{wittensl2z,mr}, related to adding Chern-Simons interactions for the gauge fields on the boundary, or to explicitly breaking the $SU(N)$ symmetry. We will not discuss these alternatives here.}. We focus primarily on the electric boundary conditions, which are more interesting in our context. In Section \ref{magsec} we describe physics with magnetic boundary conditions.

\para
The electric, or Dirichlet, boundary conditions fix those components of the gauge field that lie parallel to the boundary -- let's call them $A_i$, where $i$ is the boundary spacetime index. When we fix  $A_i=0$ on the boundary (or, in a gauge-invariant language, $F_{ij}=0$ on the boundary), the theory has an $SU(N)$ global symmetry which is generated by gauge transformations which asymptote to a non-zero constant; configurations related by such global gauge transformations are not identified, although they result in equivalent physics.
Notice that in the non-Abelian case, any choice of Dirichlet boundary condition other than  $F_{ij}=0$ explicitly breaks the $SU(N)$ symmetry. We will primarily be interested in the $SU(N)$-invariant theory.

\para
The boundary condition $F_{ij}=0$ prohibits the existence of magnetically charged states in the bulk. However, there is no problem with electrically charged states. (This situation is reversed for the Neumann boundary conditions discussed in Section \ref{magsec}).

\para
At weak coupling, the gauge  theory is deconfined and the excitations are electrically charged gluons.
The lightest state is a single-gluon state, with energy $E=2/L$ arising from the AdS curvature.  The energies of multi-gluon states receive corrections that we can compute order by order in $g^2$. However, importantly, the energy of the single-gluon state receives no corrections, since it
is fixed by the representation theory of the $SO(3,2)$ isometry group of AdS${}_4$. Just as massless spin one particles in flat space sit in a short representation, so too do gauge bosons of energy $E=2/L$ in \ads (this representation is simply the conserved current representation of the three dimensional conformal algebra $SO(3,2)$). At weak coupling there is no other multiplet that can join with this short multiplet to form a single particle state sitting in a long representation with higher energy. The energy of the gauge bosons is thus expected to remain $E=2/L$, at least for some finite range of values of $g^2$.

\para
 In the context of the AdS/CFT correspondence, after imposing Dirichlet boundary conditions  it is natural to compute the generating functional for the correlation functions of an $SU(N)$ global current $J^i$. These currents\footnote{Note that even though the four dimensional gauge theory on \ads has a global symmetry, there is no local four dimensional current, since the symmetry is gauged in the bulk.} couple to the gauge fields on the boundary and are covariantly conserved,  obeying ${\cal D}_iJ^i=0$. Denoting the boundary value of the bulk gauge field $A_i$ by $a_i$, the generating function is simply the partition function for the theory on \ads with the boundary condition $A_i |_{\partial AdS} = a_i$,
\be Z_{\rm bulk}[a_i] =  \left< \exp\left(i\int_{\partial AdS} {\rm Tr}\,(a_i(x) J^i(x))\right)\right>. \label{J}\ee
Taking derivatives of $\log(Z_{\rm bulk})$ with respect  to $a_i$, and subsequently setting  $a_i=0$,  gives the correlation functions for our theory.
Since we are not considering dynamical gravity in this section, the generating function \eqn{J} will not provide correlation functions in a well-defined three dimensional conformal field theory. In particular, the boundary theory has no stress-energy tensor. Nonetheless, we can formally use  \eqn{J} to compute correlation functions of $J$, in the family of theories parameterized by $\Lambda L$. For any value of $\Lambda L$ these theories  have the same symmetries as correlation functions of conserved currents in a CFT\footnote{Note that although  varying $\Lambda L$ gives rise to a one-parameter family of generating functions, $\Lambda L$ does not correspond to a marginal deformation in the three dimensional sense; such marginal operators are instead dual to massless scalars in the bulk. We will revisit this issue in Section \ref{turnonsec} when we discuss theories with gravity that have more honest CFT duals.}. This provides us with the two-point function which, up to contact terms, takes the form \eqn{jj}. As in the AdS/CFT correspondence, the operator $J^i$ may be associated to the single-gluon state by a state/operator map.

%
%
%
%
%
%
%

\subsubsection*{Strong Coupling}

At strong coupling, $\Lambda L \gg 1$, the gauge theory exhibits confinement at a scale $\Lambda$ which is much higher than the scale $1/L$ where the curvature of AdS becomes noticeable. The spectrum  no longer contains states charged under the  $SU(N)$ global symmetry  but instead consists of neutral glueballs with mass $\sim \Lambda$; in particular the gluon state with $E=2/L$ has disappeared.
%
Since we have argued that the single particle gluon state survived for some finite window of $\Lambda L\ll 1$, the theory must go through a sharp phase transition at a critical value of $\Lambda L$ for any $N\geq 2$. In the large $N$ limit this  phase transition further involves a jump in the behaviour of the free energy (which is just the Casimir energy for the gauge theory on AdS${}_4$),
from ${\cal O}(N^2)$ in the deconfined phase to ${\cal O}(1)$ in the confined phase.

\para
In flat space, for theories with no dynamical fields in the fundamental representation, the Wilson loop provides the canonical order parameter to differentiate between confining and deconfined phases. In AdS it is not useful because, at large distances, perimeters and areas scale in the same way \cite{cw}.
Indeed, the computation of the bulk Wilson loop is not dissimilar to the familiar Wilson loop computation in AdS/CFT, now
using the confining string in the bulk (ending on the bulk Wilson loop) rather than the fundamental string (ending on a Wilson loop on the boundary).

\para
A better order parameter is simply the existence in the spectrum of the gluon state with $E=2/L$, or the (closely related) existence of a non-trivial $SU(N)$ global symmetry acting on the spectrum. In the deconfined phase we have the operator $J$ with correlators obeying \eqn{jj}. In the confined phase it is not clear if such an operator exists; we expect changes in the boundary conditions as in \eqn{J} to affect the behaviour only at distances of order $1/\Lambda$ from the boundary, which go to zero from the perspective of a three dimensional CFT when we take the UV cutoff to infinity. Thus, to the extent that an operator $J$ may be defined in this phase, its correlators should vanish below the UV cutoff scale.
Note that this order parameter is well-defined even in theories like QCD, that have dynamical fields in
the fundamental representation of the gauge group.

\para
Finally, we can also look at a second, canonical, order parameter for confinement: the Polyakov loop. (Like the Wilson loop, this is only a good order
parameter in the absence of dynamical fundamental fields). The Polyakov loop is defined in the Euclidean theory with a periodic
time direction, corresponding to finite temperature, as the trace of the holonomy of the gauge field around the
Euclidean thermal circle. The theory with the circle has a $\ZZ_N$
global symmetry coming from gauge transformations that are only periodic around the circle up to an element
of the $\ZZ_N$ center of the $SU(N)$ gauge group, and the Polyakov loop is charged under this symmetry.
In flat space, this order parameter is usually invoked to study finite temperature
phase transitions. Here we instead wish to use the Polyakov loop at arbitrarily small temperatures as an order
parameter for the quantum phase transition that  arises when we vary $\Lambda L$. However, to use the Polyakov loop as an order parameter in AdS, we must revisit the question of boundary conditions. Instead of imposing  $A_i=0$ on the boundary (which, of course, includes $A_0=0$) we instead pick boundary conditions that preserve the $\ZZ_N$ symmetry.
This means that we choose the holonomy on the thermal circle at the boundary to have
eigenvalues that are the $N$ different $N$'th roots of unity. We claim that this choice of boundary conditions results in a vacuum with a constant Polyakov loop only at strong coupling. Indeed, in the confining phase we expect the Polyakov loop to vanish everywhere in the bulk, and these $\ZZ_N$-invariant boundary conditions are certainly appropriate. In contrast, at weak coupling these boundary conditions break $SU(N)$ gauge symmetry. Integrating out the massive gauge bosons induces a potential which forces the eigenvalues together. Furthermore, the AdS warp-factor means that this potential is strongest in the center of AdS, vanishing as we approach the boundary. The net effect is that with the $\ZZ_N$-invariant boundary conditions we expect at weak coupling to find a Polyakov loop that depends on the radial position in AdS.  In this sense, with appropriately chosen boundary conditions,  the Polyakov loop at very low temperatures can serve as a good order parameter for our quantum phase
transition.

\subsection{Mechanisms for the Phase Transition}

The main question that we would like to ask in this paper is how does the theory transition from the deconfined to the confining phase  as we change the coupling? Clearly this is a difficult question involving strong coupling dynamics, and we do not have a definitive answer. Instead, we provide a number of scenarios, of varying plausibility, that could give rise to the transition. It would be interesting to understand which of these scenarios actually occurs for specific gauge theories on AdS${}_4$.

\subsubsection*{Continuous Phase Transition}

As in other phase transitions, the deconfined vacuum at weak coupling could join continuously
to the confined vacuum at strong coupling, or there could be a jump between them at some
critical value of the coupling. We begin by analyzing the continuous case, by which we mean
that for every value of $\Lambda L$ there is an AdS-invariant vacuum state, that interpolates
between the deconfined state at weak coupling and the confined state at strong coupling. There
are two scenarios in which we can imagine this happening, but both are somewhat problematic.


\para
The first continuous scenario involves the Higgs mechanism in \ads.
As we explained above, at weak coupling the mass of the single gluon states -- corresponding to the conserved current on the boundary -- cannot vary continuously from $E=2/L$. This is because the gauge bosons sit in a short multiplet which is protected as long as the gauge symmetry survives. This, of course, is a familiar story in flat space too. In both flat space and in AdS, gauge bosons can only smoothly achieve a mass through the Higgs mechanism, eating a massless scalar field which breathes life into the longitudinal gauge mode. In order to invoke such a mechanism to explain the confinement transition, we need that as $g^2$ increases, a charged scalar state must become massless (with energy $E=3/L$) and condense, providing a mass to the gluons. At weak coupling there is no scalar state with $E < 4/L$, so clearly this can only happen once the coupling becomes of order one. At larger values of $\Lambda L$ the current two-point function would then involve a different power law than \eqn{jj}, depending on $g^2$. In AdS space there seems to be no invariant distinction between the Higgs and confinement phases, since the Wilson loop is not a good order parameter. Thus, and since the $SU(N)$ global symmetry is explicitly broken by the Higgs mechanism, it is possible that the Higgs phase smoothly joins the confinement phase at strong coupling, with the massive gluons (and all other charged states) smoothly morphing into glueballs.

\para
While such a scenario may be possible, it seems rather unlikely to us. There are many more multi-gluon states than glueball states (especially if $N$ is not small) so it is hard to imagine them joining together. Moreover, at least at weak coupling, a Higgs mechanism requires an explicit breaking of the $SU(N)$ global symmetry by the boundary conditions for the charged scalar field that acquires a vacuum expectation value. In our case the transition happens at strong coupling, so a description in terms of a scalar field in the bulk may not be appropriate, but still it seems unlikely that the $SU(N)$-invariant boundary conditions that we started from at high energies in our asymptotically free gauge theory will lead to a Higgs mechanism at low energies. (It is possible that such a Higgs mechanism could arise from $SU(N)$-breaking boundary conditions at high energies, but we are not discussing such boundary conditions here).

\para
A second continuous scenario is that
there is no Higgs mechanism, but rather, as we increase the coupling,
the two point function $\langle JJ\rangle$, which, up to a constant, obeys \eqn{jj} in the deconfined phase,  goes smoothly  to zero at some critical value of $\Lambda L$. This means that the single gluon states go to zero norm (and then decouple).
Note that the normalization of the single-gluon state is determined by the normalization of the currents that are associated with this state, which is fixed as discussed before \eqn{jj}. All other charged states should also decouple.
Note that states other than single gluon states are not in protected multiplets, and could either have a divergent energy or follow their single gluon cousins in becoming null. Usually in systems with a discrete spectrum, states cannot just disappear from the spectrum, but AdS space has infinite volume so perhaps this is not impossible (for instance, one could imagine that a state continuously
becomes non-normalizable and decouples as the coupling is varied).


\subsubsection*{Discontinuous Phase Transition}

As with any other phase transition, it is possible that the confined vacuum at strong coupling is not continuously related to the deconfined vacuum at weak coupling; because of the problems with the previous scenarios, this seems like a more likely possibility. In this scenario, as we increase the coupling, the deconfined vacuum should either cease to exist (as an AdS-invariant state\footnote{Note that there are two possible definitions of a vacuum for field theories in AdS space. One is the minimal energy configuration that obeys the boundary conditions that we impose, and the other is the minimal energy AdS-invariant configuration that obeys the boundary conditions. Since non-AdS-invariant configurations are complicated to analyze, in this paper we use the second definition. However, vacua of this type do not have to exist, and we will encounter several cases where they do not exist.}) at some value of the coupling, or it could continue to exist at strong coupling with a larger energy (by some function of $\Lambda L$) than the confined vacuum.
Similarly, the confined vacuum should either start existing (as an AdS-invariant state) at some critical coupling, or have a non-perturbatively large energy in the weak coupling limit. Vacua can stop existing, for instance, by developing tachyons (scalars below the Breitenlohner-Freedman bound $m^2=-9/4L^2$ \cite{Breitenlohner:1982jf}), or by developing runaway potentials for massless scalars (though this seems unlikely in a pure Yang-Mills theory).
It may be that there is a range of couplings in which no AdS-invariant vacuum exists, or it may be that there is a range of couplings in which both deconfined and confined AdS vacua are allowed, presumably with different energies.

\para
 There is, in fact, a plausible mechanism which could destabilize the weakly coupled deconfined vacuum as $g^2$ is increased. Consider a charged particle in AdS with mass $m\gg 1/L$, so that it can be treated effectively as a point particle. With boundary conditions $A_i=0$ on the gauge field, the boundary of AdS acts very much like an electric conductor. A point charge feels an image charge behind the boundary with opposite sign (or, for non-Abelian gauge fields, in the conjugate representation), which gives rise to an attractive Coulomb force towards the boundary. This is counterbalanced by the usual dependence of the energy on the radial position (coming from the AdS metric), which drives particles away from the boundary.
Suitably close to the boundary, we may approximate global AdS by the Poincar\'e patch with the boundary lying at $z=0$. The potential from the rest energy and from the Coulomb force then takes the form
\be V = \frac{mL}{z} - \frac{g^2}{4\pi (2z)}\label{donotrust}\ee
(the precise coefficient of the last term depends on the representation that the particle is in).
We see that for $g^2 \gg mL$ there is an instability, which can lead to particle anti-particle pairs being produced and rushing towards the boundary.
(Note that the two forces have the same form because we work in the Poincar\'e patch; in global AdS the two terms in \eqn{donotrust} do not take exactly the same form, but they still do so approximately near the boundary.)

\para
Of course, the potential \eqn{donotrust} should be taken with a grain of salt in the regime of the instability. The first term is only valid when $mL\gg 1$; the second when $g^2\ll 1$.  When we trust both terms, no instability arises. Nonetheless, it is tempting to speculate that such an instability may occur for non-Abelian theories and lead to the confinement transition in AdS\footnote{The instability we find is somewhat analogous to the instability of adjoint Wilson lines in the ${\cal N}=4$ SYM theory, analyzed in \cite{Klebanov:2006jj}. The insertion of such a Wilson line is fine at weak coupling, but leads to a spectrum which is unbounded from below at strong coupling. We thank J. Maldacena for this analogy.}. Presumably, for Abelian theories it does not occur.

\para
In all scenarios except for the second continuous scenario,
bulk confinement seems to place an upper bound on $g^2$ in \eqn{jj}, and hence a lower bound on the number of charged degrees of freedom for non-Abelian currents in a putative dual CFT$_3$ (once we couple our theories to gravity, as discussed in the next section). Another way to say this is that it seems that when the number of charged degrees of freedom in the CFT$_3$ drops below a critical value, the bulk theory undergoes confinement. The exact nature of the  bound presumably depends on the precise matter content in the bulk, since we do not expect to get confinement when there are too many charged matter fields in the bulk, nor when the theory at the AdS scale is effectively higher dimensional. We note that  a lower bound on the number of charged fields in 4d SCFTs was recently observed in \cite{poland}. For completeness, we also note that an upper bound on the number of charged degrees of freedom has also been suggested \cite{ritz}, which simply follows from the intuitive idea that they can't exceed the total number of degrees of freedom. From the bulk perspective, the fact that $g^2$ cannot be arbitrarily small sits well with the ``gravity is the weakest force" conjecture of \cite{nima}, and we are saying that in some cases it also cannot be arbitrarily large.

\subsection{Magnetic Boundary Conditions}\label{magsec}

In this subsection we describe what happens when we impose magnetic, or Neumann, boundary conditions on the $SU(N)$ gauge field. Here, we set $F_{\mu\nu}n^\nu=0$ on the boundary, where $n^\nu$ is normal to the boundary. From the perspective of the AdS/CFT correspondence, these boundary conditions gauge the $SU(N)$ global symmetry of the CFT \cite{wittensl2z,mr}, by making the boundary value of the gauge field ($a_i$ in \eqn{J}) dynamical. In an Abelian gauge theory, these boundary conditions are related to the Dirichlet boundary conditions by electric-magnetic duality. In this case the Neumann boundary conditions also exhibit a conserved global symmetry current (which is the Hodge dual of the boundary gauge field), but this is not true for non-Abelian gauge theories.

%
%
%
%
%

%

\para
In global AdS${}_4$, the Neumann boundary conditions ensure that the bulk dynamics lies under the dominion of Gauss' law, with only electrically neutral states allowed. In contrast, magnetically charged states are perfectly acceptable with Neumann boundary conditions. In this case there is no order parameter distinguishing the deconfined phase from the confined phase; in particular the single-gluon state is not part of the spectrum even at weak coupling. Also in the large $N$ limit, in both phases the free energy at low temperatures would be ${\cal O}(1)$, and at high temperatures ${\cal O}(N^2)$. The situation is similar to the case of weakly coupled Yang-Mills theory on ${\bf S}^3$, discussed in \cite{Sundborg,hagedorn}. Note that at low temperatures the gauge contribution to the free energy of the bulk theory is of order ${\cal
O}(1)$ with magnetic boundary conditions, while it is of order ${\cal O}(N^2)$ with electric
boundary conditions. At weak coupling it is easy to check this by explicit computations (similar to those of \cite{amr}); we shall consider the effects of a finite temperature further in the next subsection.

\para
With magnetic boundary conditions in place, the weak coupling dynamics appears similar to the strong coupling dynamics. Both allow only singlet states and have ${\cal O}(1)$ free energy. These simple considerations suggest that the physics may change smoothly as we vary $\Lambda L$, with the lightest $SU(N)$-singlet two-gluon states smoothly evolving into glueballs at strong coupling. Note in particular that the potential instability discussed above for the deconfined phase does not exist when we have Neumann boundary conditions, since in this case the image charge carries the same sign as the original charge, so charged states are repelled from the boundary also by the electro-static force. In fact, the confining behaviour on \ads is qualitatively similar to the weak coupling behaviour with Neumann boundary conditions\footnote{OA thanks J. Maldacena and E. Witten for discussions about this issue.}. If at strong coupling we separate an external quark and anti-quark in \ads to large distance, at least one of them must approach the boundary, where it would feel a potential coming from a confining string pulling it inside; but the $z$-dependence of such a potential is the same as that of the weak coupling potential coming from an image charge, which is the same as in \eqn{donotrust} (just with opposite sign).
In the next subsection we will look at finite temperature effects, and we will see that for high enough temperature there are nevertheless interesting differences between the weak and strong coupling phases, also with the Neumann boundary conditions.

\subsection{Finite Temperature Behaviour}
\label{finite_temp}

In the weakly coupled deconfined phase, with Dirichlet boundary conditions, nothing much happens as we change the temperature. In particular, in the large $N$ limit the free energy is ${\cal O}(N^2)$ at all temperatures. This is not true with Neumann boundary conditions. In this case, a similar computation to those of \cite{hagedorn,amr} shows that at weak coupling the free energy is ${\cal O}(1)$ at low temperatures, while the high temperature result is insensitive to the boundary conditions and is ${\cal O}(N^2)$. In the large $N$ limit there is a sharp deconfinement transition between these phases, which at zero coupling happens at $T = 1 / \ln(3) L$. Note that this is a large $N$ phase transition as we change the temperature, which is not directly related to the finite $N$ transitions as we change parameters that we discussed above.

\para
One can also ask what becomes of the bulk confined phase when it is heated. This phase has a mass gap of order $\Lambda$, but we can ask what happens when we heat to a temperature above $T\sim \Lambda$.
 At this point, one might expect a small bubble of deconfined plasma to appear in the bulk (in this discussion we assume that the deconfinement phase transition in flat space is of first order, as in pure $SU(N)$ gauge theories with $N \geq 3$).
The Euclidean \ads metric in global coordinates is given by
\be \label{adsmetric}
ds^2 = L^2 \left(\cosh^2(\rho) d\tau^2 + d\rho^2 + \sinh^2(\rho) d\Omega_2^2\right). \ee
We periodically identify the imaginary time direction, $\tau \in [0,\beta)$. In the AdS/CFT context this corresponds to a temperature
for the dual CFT given by
\be \label{dualT} T = \frac{1}{\beta L}.\ee
However, the local temperature in the bulk depends on the radial coordinate $\rho$. It is given
by
\be T_{\rm local}(\rho) = \frac{1}{\beta L}\frac{1}{\cosh(\rho)} \leq T,\ee
and reaches the temperature \eqn{dualT} only in the center of AdS, $\rho=0$. For temperatures larger than the flat space deconfinement temperature $T_{crit}$ (which is of order $\Lambda$), we thus expect a ``plasma-ball'' to form in the center of AdS space whenever $T > T_{crit}$, with a bubble of deconfined plasma filling the region $\rho < \rho_\star$, where $T_{\rm local}(\rho_\star) \simeq T_{crit}$.
For $T \gg T_{crit}$ the size of the bubble is given by $e^{\rho_\star} \sim 2 T/T_{crit}$.

\para
With the Neumann boundary conditions, the appearance of such bubbles at high temperatures is one of the features which distinguishes the weak coupling (deconfined) phase from the strong coupling (confined) phase.

\section{Turning on Gravity}\label{turnonsec}

In this section we describe how some of the above results change in the presence of dynamical (weakly coupled) gravity in the bulk. In this case, the AdS/CFT correspondence implies that the bulk dynamics gives the partition function for a large $\hat N$ three dimensional conformal field theory. Note that this $\hat N$, which controls the gravitational coupling in the bulk (it is related to $L M_p$), has nothing to do with the rank $N$ of the gauge theory we have been discussing above. For weak gravity $\hat N \gg N$, and the bulk gauge theory includes just a small fraction of the dynamical states of the large $\hat N$ three dimensional CFT. The weak gravity assumption means in particular that the Planck scale $M_p$ is much larger than the dynamical scale $\Lambda$.

\para
With electric boundary conditions in place, when $\Lambda L\ll 1$ the state-operator map ensures that the dual CFT on the sphere has states $J^i|0\rangle$ with energy $2/L$, sitting in an adjoint multiplet of the $SU(N)$ global symmetry. These correspond to the gluons in the bulk carrying colour charge which we discussed above.  In contrast, at strong coupling, $\Lambda L \gg 1$, the gauge theory exhibits confinement and  the spectrum of asymptotic states consists of
glueballs with mass $\sim \Lambda$. By the usual AdS/CFT dictionary, these correspond to operators
in the dual CFT of dimension  $\Delta \sim \Lambda L \gg 1$. So, the sector of the large $\hat N$ CFT associated with the bulk $SU(N)$ theory must develop a large gap in its spectrum of dimensions, and all states charged under the $SU(N)$ global symmetry must disappear.

\para
It is tempting to say that the dual CFT${}_3$ has exhibited confinement for the non-Abelian global $SU(N)$ symmetry. However, confinement for global symmetries is unfamiliar and this does not appear to be the correct interpretation.  In general in AdS/CFT, changing bulk parameters does not leave us in the same theory.
Rather we are moving through different dual CFTs, presumably with different
Hilbert spaces.
It is likely that in any string embedding\footnote{This statement refers to embeddings with four supercharges or less, which is the interesting case for four dimensional confining gauge theories.}, $\Lambda L$ is not a
continuous parameter but instead takes values in a discretuum, and theories with values of $\Lambda L$ which are close to each other need not necessarily lie close by in the landscape of string vacua. For this reason, it is not clear that one can construct a one-parameter family of closely related CFTs which have continuously varying $\Lambda L$, such that they undergo a confining phase transition of the type described in the previous section. More precisely, it is clear that changing $\Lambda L$ is not an exactly marginal deformation in the dual CFT${}_3$, since generically there is no massless scalar field in the bulk whose vacuum expectation value controls this parameter. It is possible that there is a discrete parameter (like a Chern-Simons coupling) in the CFT${}_3$ that controls this parameter and which becomes continuous in the large $\hat N$ limit, and then we claim that we have a phase transition as a function of this approximately-continuous parameter, but this does not have to be the case, and it may depend on the precise string theory construction.

\para
In any case, one could ask what is the signature of bulk confinement (at large $\Lambda L$) in the dual CFT${}_3$. Namely, how can a CFT${}_3$ observer in the strong coupling phase detect the presence of the confined bulk gauge symmetry. It seems that this clean question does not have a simple answer. One way to proceed when the bulk gauge theory is asymptotically free is to
look at high energy glueball scattering in the bulk, at energies that are much higher than $\Lambda$ but much smaller than $M_p$. This is related  to 4-point CFT$_3$ correlation functions of the operators dual to glueballs. (See, for example, \cite{mellin1,mellin2} for recent progress in relating CFT${}_3$ correlation functions to bulk scattering amplitudes). These correlation functions will have a specific form when these glueballs arise from an asymptotically free gauge theory, and in the high energy limit one should be able to see how many ``partons'' each confined state is made of. Similar remarks have been made in \cite{loga}. We will discuss other possible signatures in the next subsection.

\subsection{Probing the (De)-Confinement Transition}\label{flows}

Although varying the bulk parameter $\Lambda L$ cannot be thought of as a strict phase transition in the dual CFT, one could nonetheless ask if there is some deformation of this theory that would lead to a confinement/deconfinement phase transition.

\para
A holographic renormalization group flow offers a simple example of such a transition. One could imagine flowing between two CFT$_3$'s, both described by \ads spacetimes with an $SU(N)$ gauge symmetry. The radius of AdS for the ultra-violet (UV) CFT$_3$ is $L_1$; the radius of AdS for the IR CFT$_3$ is $L_2 < L_1$. If the gauge coupling is independent of this flow, one could arrange that $\Lambda L_1 \gg 1$ while $\Lambda L_2\ll 1$, so that the bulk gauge theory flows from the confining phase in the UV to the deconfined phase in the IR. (Note that this holographic confinement transition occurs in the opposite manner to the usual renormalization group flows of the bulk gauge theory, which are asymptotically free in the UV; this is a manifestation of the usual UV-IR relation.) In this setting, the deconfined phase in the IR geometry has the interpretation of an emergent $SU(N)$ global symmetry in the $CFT_3$.

\para
It appears to be  somewhat harder to orchestrate a situation in which the deconfined phase exists in the UV, flowing to a confined phase in the IR. Presumably it can be achieved in solutions where the bulk gauge coupling (perhaps related to a dilaton field) becomes larger along the flow. In such a case the global $SU(N)$ symmetry would appear to decouple at low energies, as if it couples only to massive fields, but here its effects would still be present (just hidden by confinement) in the IR.

\para
Naively, another way to detect the presence of the confined gauge group is by heating the theory. As discussed in the previous section, in the absence of dynamical gravity, one can make a bubble of deconfined phase inside the confined phase by heating to high enough temperatures. However, the presence of dynamical gravity changes this conclusion. The Hawking-Page phase transition \cite{Hawking:1982dh,Witten:1998zw} to the black hole always occurs at $T_{\rm HP}\sim 1/L$. In the confining phase, $\Lambda L \gg 1$, ensuring that $T_{\rm HP} \ll T_{crit} \sim \Lambda$. The formation of the black hole thus occurs before the creation of the plasma-ball, or in other words, the plasma-ball is always hidden behind a horizon. More precisely, this is true in the canonical ensemble; in the micro-canonical ensemble there could still be approximately stable plasma-balls when $\Lambda L \gg 1$.

\subsection{String Theory Constructions}

In the AdS/CFT correspondence, one usually assumes for simplicity that the bulk theory is weakly coupled. Indeed, often we consider this correspondence for a large ${\hat N}$ limit of some gauge theory which, in the bulk, has a small string coupling $g_s \sim 1/{\hat N}$. However, this does not prevent gauge couplings in the bulk from becoming large and confining, if $\hat N$ is large but not infinite, and if we have some asymptotically free gauge theory in the bulk. Such a theory could come, for instance, from space-filling D-branes or from Kaluza-Klein reductions on spheres.  The only requirement is that the AdS radius $L$ must be large enough so that $\Lambda L \gg 1$. When $\Lambda$ arises by dimensional transmutation in a weakly coupled 4d gauge theory, it is exponentially small compared to the typical scales of the string theory, so this requires an exponential fine-tuning of the cosmological constant. However, such a fine-tuning is necessary also for the description of our four-dimensional world, and it is widely believed that many vacua obeying such fine-tunings exist in the landscape of all string theory vacua. Gauge theories on \ads could also arise in string theory from branes filling an $AdS_4$ subspace of a higher dimensional AdS space (say $AdS_5$), as in \cite{Karch:2000ct}; in this case they would be decoupled from four dimensional gravity, but would still couple to a higher dimensional gravitational theory.


\para
One simple model in which such a fine-tuning can be arranged is a KKLT compactification \cite{kklt} in which we put $N$ anti-D3-branes into a Klebanov-Strassler throat of some Calabi-Yau compactification \cite{Klebanov:2000hb,Giddings:2001yu}. Due to the lack of supersymmetry we expect the scalars and fermions on these anti-D3-branes to become massive\footnote{Except for some singlet scalars (fermions) which are Nambu-Goldstone bosons (fermions) for the global symmetries (supersymmetries) broken by the anti-D3-branes \cite{Aharony:2005ez}.} so that the D3-branes harbour a pure $SU(N)$ gauge theory at low energies with some strong coupling scale $\Lambda$. The claim of \cite{kklt} is that, in such backgrounds, the cosmological constant can be fine-tuned to be small and positive through the use of discrete fluxes. However, the same arguments allow for fine-tuning a small negative cosmological constant such that $\Lambda L\gg 1$, resulting in a confining gauge theory in AdS.

\para
More controllable examples come from supersymmetric vacua. For example, it was argued in \cite{adssmall} that one may construct supersymmetric AdS${}_4$ vacua in which the sizes $R$ of internal cycles are parametrically smaller than the AdS scale $L$. Wrapping space-filling D-branes around the internal cycles results in a gauge theory on \ads with UV cut-off set by the scale $1/R$ and, correspondingly, $R\Lambda\ll 1$ at weak coupling. Once again, tuning $L$ relative to $R$ may allow us to realize the strong coupling regime $\Lambda L \gg 1$.

\para
Note that one could worry that the cosmological constant associated with the confining theory itself would already be too big for the fine-tuning described above to occur. However, this does not seem to be the case. In the absence of supersymmetry, we expect the vacuum energy density of a confining theory to be of order $\Lambda^4$ (with some sign). If this (together with other similar contributions) has a negative sign, we can identify it with the AdS cosmological constant $-M_p^2/L^2$, and we get $\Lambda L \sim M_p / \Lambda$, which is always much larger than one in our constructions.

\para
With supersymmetry, the cosmological constant $-M_p^2/L^2$ induced by the strong coupling dynamics is even smaller. In supersymmetric examples one can have situations where the full AdS cosmological constant comes from the non-Abelian gauge dynamics.  Both in type IIA with D6-branes,
and in F-theory with D7-branes, one can engineer ${\cal N}=1$ $SU(N)$ super Yang-Mills. The cosmological constant in a supersymmetric vacuum is $V_0\sim - |W|^2/M_p^2$
 (up to some irrelevant factor of the K\"ahler potential). If we assume that we have in the bulk an ${\cal N}=1$ super Yang-Mills theory with $\Lambda L \gg 1$, the superpotential $W$ is given by the gaugino
condensate, $|W|\sim \Lambda^3$. Assuming that this is the only contribution to the cosmological constant, this results in an AdS scale $\Lambda L \sim (M_p/\Lambda)^2\gg 1$, so our assumption that $\Lambda L \gg 1$ is self-consistent.

\para
Unfortunately, as far as we know, the dual three dimensional CFT is not understood for any of the examples of this type in which $\Lambda L \gg 1$. It would be interesting to find this CFT in some examples, in order to make our questions about identifying the confining gauge group in this CFT${}_3$ more concrete.


\section{Supersymmetric Theories in Anti-de Sitter Space}\label{susysec}

The study of supersymmetric field theories in Minkowski space has been an immensely useful tool to understand strong coupling dynamics. For this reason, in this section we discuss a number of aspects of supersymmetric Yang-Mills theories in AdS${}_4$, hoping that they will also be useful for understanding confinement in AdS. However, we will see that while the supersymmetric theories raise several interesting new issues about strongly coupled dynamics in AdS${}_4$, they do not really help  understand confinement as far as we can see.

\para
${\cal N}=1$ supersymmetric theories in AdS${}_4$ were constructed long ago in \cite{Zumino:1977av,Ivanov:1980vb,Sakai:1984nc,Burgess:1984rz,Burgess:1984ti,gary}, and recently revisited in \cite{allan,fs}.
The low-energy bosonic action for a single chiral superfield $\Phi$ (the generalization to several chiral superfields is straightforward) is written, as in flat space, in terms of a holomorphic superpotential $W(\Phi)$ and a K\"ahler potential $K(\Phi,\Phi^\dagger)$. It takes the form (using the notation $\Phi$ also for the bottom component of the chiral superfield)
%
%
%
\be {\cal L} =  K_{\Phi\Phi^\dagger} \partial_{\mu} \Phi \partial^{\mu} \Phi^\dagger -
K^{\Phi\Phi^\dagger} \left|\frac{\partial W}{\partial \Phi} + \frac{1}{L} \frac{\partial K}{\partial \Phi}\right|^2 +
\frac{3}{L} \left(W + W^\dagger + \frac{K}{L}\right),
\label{susyact}
\ee
where $K_{\Phi \Phi^\dagger}$ is the second derivative of the K\"ahler potential, and $K^{\Phi \Phi^\dagger}$ is its inverse. Note that even if our theory in flat space has some $U(1)_R$ symmetry, this is explicitly broken by \eqn{susyact}, related to the fact that one cannot put chiral fermions on $AdS_4$ \cite{Allen:1986qj,ben,Porrati:2009dy}.
In addition, fields in the same supersymmetry multiplet that had the same mass in flat space no longer have equal masses in $AdS_4$.
%
%

\para
Notice that for finite $L$ \eqn{susyact} is no longer invariant under the naive K\"ahler
transformations $K\to K + \Gamma(\Phi)+ \Gamma^\dagger(\Phi^\dagger)$ for holomorphic $\Gamma$. Instead the K\"ahler transformation must be supplemented by a shift of the superpotential,  $W \to W -\Gamma(\Phi) / L$.  This means that on AdS, there is no invariant distinction between the superpotential and K\"ahler potential, and the resulting loss of holomorphic constraints means that we lose much of the power of supersymmetry in constraining the dynamics. This is not surprising since the ${\cal N}=1$ supersymmetry algebra on \ads is equivalent to the three dimensional ${\cal N}=1$ superconformal algebra, and 3d ${\cal N}=1$ theories do not have any natural holomorphic objects.
This loss of control can be seen, for example, in the condition for
supersymmetric vacua, which is now\footnote{We discuss here only vacua in which the scalar fields are constant in AdS space.}
\be \frac{\partial W}{\partial \Phi}  + \frac{1}{L}\frac{\partial K}{\partial \Phi} = 0.\label{vacua}\ee
As in flat space, solutions to \eqn{vacua}
are always extrema of the scalar potential. However, since \eqn{vacua} is not holomorphic, it is not guaranteed to yield a fixed number of solutions as parameters in $W$ and $K$ are varied. In other words, there is no Witten index in AdS${}_4$. The best we can do is to view \eqn{vacua} as two real equations in two real variables. The resulting vacua may then come and go in pairs as we vary parameters (by having solutions to the real equations move into the complex plane); this happens even in the simplest example of a Wess-Zumino model on \ads with $W=\frac{1}{2}m \Phi^2 + \frac{1}{6} g \Phi^3$ \cite{Ivanov:1980vb}. Even though the number of supersymmetric vacua can vary, this argument tells us that, counting with degeneracies, the number of supersymmetric vacua modulo two in the theory must be independent of any continuous parameters; it must be either even or odd. More precisely, this is true assuming that no vacua go off to infinity or become non-normalizable states, as we vary parameters.



\subsection{${\cal N}=1$ Super Yang-Mills}

Let us focus on a specific example: $SU(N)$, ${\cal N}=1$ super Yang-Mills (SYM) theory on AdS${}_4$. At weak coupling, $\Lambda L\ll 1$, the theory is well described by the classical action governing the massless gauge field and massless gaugino. (There is in fact a subtlety here: the gaugino appears to gain a local UV mass term at one-loop, but this is precisely canceled by a one-loop contribution arising from the boundary \cite{ben}). There is a one-parameter family of boundary conditions for the fermion, each of which explicitly breaks the $U(1)_R$ symmetry of the theory. Classically all choices are equal, but due to the anomaly, only choices related by a $\ZZ_{2N}$ phase are equivalent in the quantum theory. Fixing the boundary condition fixes the action of the supersymmetry generators in AdS. For each choice, fixing Dirichlet boundary conditions for the gauge field, there is then a unique supersymmetric vacuum with a global $SU(N)$ symmetry current supermultiplet on the boundary.

\para
At strong coupling, the situation is very different. It is not obvious what are the lightest degrees of freedom, but in flat space one can analyze the supersymmetric vacua by writing the action in terms of an effective degree of freedom which is the chiral superfield $S \sim {\rm tr}(W_{\alpha}^2)$, whose lowest component is the gaugino-bilinear condensate \cite{vy}. When $\Lambda L \gg 1$, we expect this to be the case also in AdS${}_4$. In this limit the flat space superpotential \cite{vy}
\be
W_{VY} = S (\log(S^N/\Lambda^{3N})-N)
\label{sympot}
 \ee
 is a good approximation to the effective superpotential on \ads, with small corrections at order $1/L$ (including the second term in \eqn{vacua}), and this leads (by solving \eqn{vacua}) to $N$ supersymmetric vacua, close (for large $\Lambda L$) to their positions in flat space. In each of these vacua, the expectation value of $S$ -- and hence the boundary condition on $S$ -- differs. Unlike in flat space, the vacua are not degenerate. Nonetheless, each vacuum preserves the same supersymmetry and, in this sense, they should be thought of as being vacua of the same SYM theory, descending from the single weakly coupled vacuum discussed above. (In the AdS/CFT context, after coupling our theory to gravity, each such vacuum would be a different $CFT_3$, and the relation between these $CFT_3$'s is not clear.)

\para
As discussed above, it is not surprising that we find a different number of supersymmetric vacua, $N$, at strong coupling compared to weak coupling. However,
for $N$ even, the number of vacua has changed by a odd amount, in conflict with our general statement above. What is the resolution to this? We believe the resolution involves the subtleties of the superpotential \eqn{sympot}. This superpotential involves the logarithm function which has branch cuts, and one must interpret these branch cuts carefully already in flat space  in order to see that the effective action is invariant under the $\ZZ_{2N}$ symmetry of the theory. In particular, already in flat space one cannot see all the supersymmetric vacua by staying in the same branch, since there is only one zero of $W'$ for a specific choice of branch for the logarithm. It was suggested in \cite{Kogan:1997dt}  that one should allow all different branches for the logarithm but that, in each region in $S$-space, one branch which has the minimum value of the scalar potential (proportional to $|W'|^2$) dominates (for infinite volume) over the others. There are cusps in the effective potential on the boundary between two such regions, and near these cusps the effective action using the variable $S$ breaks down (in the sense that the massive fields that have been integrated out take different values on the two sides of the cusp, so they cannot be ignored). In particular, going from one region to the other is not done just by changing field variables in the low-energy effective action. Our analysis saying that the number of vacua changes in pairs depended on assuming that the condition for a supersymmetric vacuum \eqn{vacua} changes smoothly as we change the parameters, going from weak to strong coupling (even if we change our field variables when we do this). But at strong coupling this condition takes different values in the different regions, and there is no single superpotential that captures all $N$ of the minima. As we go from weak to strong coupling, we land in one of the $N$ different regions of $S$-space, where there is a unique choice of the branch of the logarithm that dominates. In each such region the number of solutions to the equations for supersymmetric vacua \eqn{vacua} is one both at weak and at strong coupling. Since there is no superpotential that captures all $N$ vacua simultaneously, there is no contradiction with our general arguments.

\para
Of course, we still have a confinement phase transition of the type discussed in section \ref{basicsec}, as we go from weak to strong coupling. Note that the vacuum expectation value of $S$ is non-zero already at weak coupling \cite{Allen:1986qj,ben,Bolognesi:2011un}, with $|\langle S \rangle| \sim 1/L^3$, so it does not serve as an extra order parameter in these theories. In our discussion of non-supersymmetric vacua we mentioned that these could appear/disappear by having scalar fields in the bulk become tachyonic, below the Breitenlohner-Freedman bound $M^2 L^2 = -9/4$. In supersymmetric theories this cannot happen, but there are other possible ways for vacua to destabilize which are not ruled out by supersymmetry. It would be interesting to understand which of our scenarios for the confinement phase transition occurs in the ${\cal N}=1$ SYM theory on \ads.

\subsection{${\cal N}=2$ Super Yang-Mills}
\label{ntwosym}

Moving on to theories with extended supersymmetry, let us focus for simplicity on the pure ${\cal N}=2$ SYM theory with no superpotential.
The familiar Coulomb branch of ${\cal N}=2$ super Yang-Mills in flat space is lifted in AdS; the adjoint scalar $\phi$ acquires a negative mass squared (already classically), above the Breitenlohner-Freedman bound, due to the K\"ahler potential term in \eqn{susyact}. Instead, at weak coupling the theory again has a unique vacuum state, with the adjoint scalar $\phi=0$. The boundary conditions on the fermions break the $SU(2)_R$ symmetry to $U(1)_R$; this is the $U(1)_R$ symmetry which is part of the 3d ${\cal N}=2$ superconformal algebra which 4d ${\cal N}=2$ theories on \ads satisfy (in our theory, which is decoupled from gravity, this is a continuous global symmetry; upon coupling to gravity it becomes a gauge symmetry in AdS${}_4$)\footnote{We assume that the boundary conditions preserve as much supersymmetry as possible.}.

\para
At strong coupling $\Lambda L\gg 1$, the low-energy dynamics is captured by the famous Seiberg-Witten solution \cite{Seiberg:1994rs}. For simplicity, we focus on $SU(2)$ super Yang-Mills, for which the solution is parameterized by  $U={\rm Tr}(\phi^2)$. Since 3d ${\cal N}=2$ theories do have holomorphic quantities, it is possible that the prepotential can be computed exactly for any value of $\Lambda L$ and, in fact, one can argue that it is probably independent of $\Lambda L$ (since the latter sits in a background hypermultiplet). In any case, here we will just discuss the strong coupling limit (of course, for small $\Lambda L$ the Seiberg-Witten description breaks down since there are different light degrees of freedom). Solving the supersymmetry condition \eqn{vacua}, we find several vacua. Away from the singular points in the $U$-plane, the superpotential vanishes so \eqn{vacua} is simply $\partial K / \partial U = 0$, with $K$ the K\"ahler potential that is derived from the Seiberg-Witten prepotential \cite{Seiberg:1994rs}. This K\"ahler potential is invariant under $U \to -U$, so there is clearly one solution at $U=0$. We did not find any additional solutions, but if they exist it is clear that they come in pairs. In addition, there are two special points at $U = \pm \Lambda^2$, where additional degrees of freedom, monopoles or dyons, become massless. Focusing on the point $U=\Lambda^2$,  we need to add an extra monopole hypermultiplet, containing chiral multiplets $M$ and $\tilde M$, and the superpotential is $W \sim (U-\Lambda^2) M {\tilde M}$. The K\"ahler potential for $M$ and ${\tilde M}$ near this point is approximately canonical, while $\partial K / \partial U$ vanishes at $U=\Lambda^2$, so there is (for large $\Lambda L$) a supersymmetric vacuum with $U = \Lambda^2$, $M={\tilde M}=0$. A similar supersymmetric vacuum exists at $U=-\Lambda^2$.

\para
In all these vacua, as in flat space, the bulk gauge symmetry is not confined, but rather it is Higgsed from $SU(2)$ to $U(1)$. Thus, these vacua make sense only with Dirichlet boundary conditions for the gauge field, which allow for this Higgsing\footnote{As in the AdS/CFT correspondence, this Higgsing involves an explicit breaking of the $SU(2)$ global symmetry to $U(1)$.}. If we instead tried to impose  Neumann boundary conditions, an AdS-invariant supersymmetric vacuum  would necessarily involve  an explicit breaking of the $SU(2)$ gauge symmetry on the boundary, which is problematic. It is possible that if one insists on Neumann boundary conditions, the theory undergoes a ``renormalization group'' flow to Dirichlet boundary conditions, of the type discussed in section \ref{flows}.

\para
While the monopole and dyon vacua are both minima of the scalar potential, the vacuum at $U=0$ is a (stable) maximum. One can construct a supersymmetric solution in AdS${}_4$, flowing from the $U=0$ vacuum near the boundary to one of the other vacua in the interior. Note that the number of supersymmetric vacua at strong coupling is the same modulo two as the number at weak coupling, consistent with our general arguments above.

\para
In flat space, the theory undergoes confinement when a small mass deformation is added, prompting the monopole (or dyon) to condense. In AdS, it is not so simple. First, when we add a small mass deformation $W = m U$, there is still a supersymmetric vacuum near $U=0$ which does not exhibit any condensation, in which the two terms in \eqn{vacua} cancel each other. But the situation is more interesting near the other two vacua, where a naive analysis of \eqn{vacua} suggests supersymmetric vacua in which (as in flat space) monopole operators $M$ and ${\tilde M}$ condense. However, since we are discussing the case with Dirichlet boundary conditions, the magnetic $U(1)$ symmetry is effectively gauged, and inducing an AdS-invariant expectation value for, say, the bulk monopole operator $M$ explicitly breaks the magnetic $U(1)$ gauge symmetry on the boundary, which is problematic. Moreover, if there was a constant monopole condensate in AdS${}_4$, the magnetic $U(1)$ gauge field would acquire a mass, but the Neumann boundary conditions which it satisfies are not consistent for massive gauge fields.

\para
Thus, it seems likely that there is no AdS-invariant supersymmetric vacuum near the monopole and dyon points after the mass deformation, but there could be other vacua in which the fields depend on the radial coordinate (in particular, one may have configurations in which $M$ and ${\tilde M}$ are close to their flat-space values, except near the boundary). It seems clear that understanding confinement in \ads via monopole condensation is more subtle than in flat space, and it would be interesting to understand this better (in the mass-deformed ${\cal N}=2$ theory or in other examples).

\subsection{${\cal N}=4$ Super Yang-Mills}

The case of ${\cal N}=4$ super Yang-Mills on \ads was discussed in detail recently in \cite{amr,Aharony:2011yc}. For this reason, here we  mention only a few brief comments.

\para
The ${\cal N}=4$ theory has (for any value of the coupling constant) many different supersymmetry-preserving vacua in \ads. Each vacuum obeys different boundary conditions. Since the ${\cal N}=4$ theory is conformal, and \ads is conformally equivalent to a half-line, these boundary conditions are in one-to-one correspondence with the half-supersymmetry-preserving boundary conditions for the ${\cal N}=4$ theory on a half-line, that were classified in \cite{gw}. Dirichlet boundary conditions generally involve giving a non-zero constant vacuum expectation value to three of the six adjoint scalar fields of the theory, such that they sit in some $N$-dimensional representation of $SU(2)$. For instance, we can consider the $N$-dimensional irreducible representation, in which case the $SU(N)$ global symmetry is completely broken by the boundary conditions, corresponding to a Higgs mechanism in the bulk of \ads. There is also the trivial representation, with no (classical) expectation value, and Neumann boundary conditions are also possible (with no scalar VEVs).

\para
At weak coupling the ${\cal N}=4$ theory certainly does not confine, but we claim that in some \ads vacua at strong coupling it does.
There are two reliable methods to analyze what happens at strong coupling. For $g_{YM}^2 \gg 1$ we can use S-duality, and for large $N$ and large $g_{YM}^2 N$ in the 't Hooft limit we can use the AdS/CFT correspondence (note that we now use this correspondence to relate our 4d theory on \ads to a 5d theory with $AdS_4$ boundary; the relevantb gravitational solutions were analyzed in \cite{amr,Aharony:2011yc}).
S-duality exchanges the electric and magnetic fields, so one may guess that starting with the Higgs vacuum described above, and going to strong coupling, would give a vacuum with a magnetic condensate in the weakly coupled S-dual description, which should be confining. When the S-dual theory is at large $N$ and has large 't Hooft coupling ${\tilde g}_{YM}^2 N$, this can be confirmed using the AdS/CFT correspondence, using the solutions of \cite{Aharony:2011yc}. In these solutions the Higgs vacuum comes from the near-horizon limit of D3-branes ending on a single D5-brane, and the dual gravity solution also contains this D5-brane (which gives screening of Wilson lines), while the S-dual gravity solution (corresponding to D3-branes ending on a single NS5-brane) contains a NS5-brane, and exhibits confinement (in the sense of an area law for Wilson lines) and screening of 't Hooft loops. In the strong 't Hooft coupling limit where a gravity approximation is valid, one can compute the confining properties of the theory (such as the tension of the confining string) reliably using the gravitational description\footnote{Confining theories which have holographic dual descriptions, such as \cite{Witten:1998zw,Polchinski:2000uf,Klebanov:2000hb,Maldacena:2000yy}, can also be studied on \ads using generalizations of their flat-space gravity duals. However, all these theories are not asymptotically free, so their behavior at small $\Lambda L$ is different from the one discussed in this paper, and they do not exhibit phase transitions of the type discussed in section \ref{basicsec}.}. However, since the S-duality exchanges
Dirichlet and Neumann boundary conditions (though its full action on the supersymmetric boundary conditions is much more complicated \cite{gw}), the confining theory we find is for Neumann boundary conditions. With these boundary conditions, as discussed above, there is no order parameter distinguishing the confined and deconfined phases (which is good, since with so much supersymmetry one should not have any phase transitions as a function of the coupling constant). Moreover, as discussed in section \ref{basicsec}, the confining behaviour on \ads is qualitatively similar to the weak coupling behaviour with Neumann boundary conditions. So, it is likely that as we increase the coupling in this case, the weak coupling deconfined behaviour smoothly goes over to confining behaviour at strong coupling, confirming our general arguments in section \ref{magsec}. It would be interesting to analyze these theories at finite temperature, to see if they exhibit plasma-balls of the type discussed in section \ref{finite_temp}.

\section*{Acknowledgements}

During the 2.5 years that we have been working on this project, we have discussed it with many people, and we would like to thank all of them. In particular we would like to thank A. Cherman, N. Dorey, G. Festuccia, O. Ganor, D. Harlow, S. Kachru, D. Kutasov, X. Liu, S. Minwalla, J. Penedones, E. Sharpe, M. Strassler, E. Witten, and especially Z. Komargodski, J. Louis, J. Maldacena and N. Seiberg for useful discussions.
OA would like to thank the LMS-EPSRC Durham Symposium on ``Non-perturbative techniques in field theory'' in July 2010, the ``Third Indian-Israeli meeting on String Theory : Holography and its Applications'' in February 2012, the Caltech conference on ``${\cal N}=4$ SYM theory, 35 years after'' in March 2012, and the Institute for Advanced Study for inviting him to present preliminary versions of these results. OA would like to thank Durham University, Imperial College, Queen Mary University, the Technion and Tel-Aviv University for hospitality during the course of this project. DT would like to thank the Weizmann Institute (twice), as well as Tel-Aviv University and the Technion for their generous hospitality. OA is the Samuel Sebba Professorial Chair of Pure and Applied Physics, and he is supported in part by a grant from the Rosa and Emilio Segre Research Award. The work of OA, MB and SY was supported in part by an Israel Science Foundation center for excellence grant, by the Israel--U.S.~Binational Science Foundation, by the German-Israeli Foundation (GIF) for Scientific Research and Development, and by the Minerva foundation with funding from the Federal German Ministry for Education and Research. OA gratefully acknowledges support from an IBM Einstein Fellowship at the Institute for Advanced Study.
DT is supported by STFC and by the  ERC STG grant 279943, ``Strongly Coupled Systems".

\appendix

\section{Instantons in Anti-de Sitter Space}

In this paper we focused on strongly coupled dynamics associated with confinement, but we can also discuss other interesting non-perturbative effects in gauge theories on AdS space.
One useful tool to obtain information about the non-perturbative dynamics of gauge theories in flat space is instanton computations. The discussion of instantons in \ads introduces some new features which are not present in flat space (as discussed already in \cite{cw}). Naively, since \ads is conformally related to flat space, instanton solutions on \ads are the same as in flat space. However,  a general instanton solution, when mapped to AdS$_4$, will not obey the prescribed boundary conditions on the boundary of \ads (and, moreover, part of the instanton would be outside the boundary, so the total instanton number on \ads would be fractional).

\para

The discussion of instantons on \ads thus depends strongly on the boundary conditions. We will focus here on the case of Dirichlet boundary conditions, which was the most interesting case also for confinement. We require $A_{||}=0$ on the boundary. But this ensures that the instanton number $\int_{AdS_4} {\rm tr}(F \wedge F)$, which  can also be written as $\int_{\partial AdS_4} {\rm tr}(A\wedge F+ \frac{2}{3} A^3)$, is identically zero. So, Dirichlet boundary conditions do not allow instantons (though they can allow instanton-anti-instanton pairs), and it seems that the $\theta$ angle does not play any role.

\para
However, we should be more careful. There are two separate issues. The discussion above assumed that the boundary of Euclidean \ads is taken to be $S^3$. However, if we are interested in the analytic continuation of Lorentzian physics on global AdS$_4$, it is more natural to take
Euclidean \ads to have the metric \eqn{adsmetric}, in which the boundary is $S^2 \times \RR$.
%
%
Our spatial slices are then three dimensional disks, with the requirement that $A_{||}=0$ on their boundary. The allowed gauge transformations are now mappings $g$ from the gauge group $G$ to the 3-disk, with the boundary condition $g=I$ at the boundary. This is the same as in the usual discussion of instantons in flat space, so we obtain the usual conclusion about the existence of $n$-vacua. Instantons now tunnel between the different $n$-vacua which, after diagonalization, are re-expressed as the usual   $\theta$-vacua. Note that the relevant instantons now have flux escaping only to $\tau \rightarrow \pm \infty$ and not to spatial infinity, where we impose the boundary conditions $F_{||}=0$ \footnote{From the point of view of the ``standard'' Euclidean \ads whose boundary is $S^3$, we allow singular configurations at two points on the $S^3$, and this allows instantons to appear, and to play similar roles as in flat space.}. For example, in computations relevant for ${\cal N}=2$ SYM theory in global AdS$_4$, the $\theta$ angle would appear in the Seiberg-Witten prepotential which we discussed in section \ref{ntwosym}.

\para
In the ``standard'' Euclidean \ads we do not have such a sum over instantons but, nevertheless, our conclusion that the $\theta$ angle plays no role was too hasty. For example, in the Seiberg-Witten theory discussed in section \ref{ntwosym}, we claimed that the expectation values of scalar fields in the supersymmetric vacua depended on $\Lambda$ (whose phase is given by the $\theta$ angle), and this should be visible also in Euclidean computations. The point is that the theta angle in the bulk can be written (by Stokes' theorem) \cite{wittensl2z} as the coefficient of a Chern-Simons term for the background gauge field coupling to the global symmetry current $J^i$. Such a term does not affect correlators of currents at separated points, but it does affect contact terms (for instance, in
the 2-point functions of currents), as recently discussed in \cite{Closset:2012vg,Closset:2012vp}. Indeed, it was stated in these references that the value of the contact term in the 2-point function of currents modulo one (in some normalization) is physical, and in our theories this contact term is precisely equal to the $\theta$ angle divided by $2\pi$. So, even in the ``standard'' Euclidean AdS$_4$, the effect of the $\theta$ angle appears through this contact term.

\para
This does not yet explain how the position of supersymmetric vacua changes when we change the $\theta$ angle, as Seiberg-Witten theory implies. To understand this we note that in Seiberg-Witten theory with supersymmetric boundary conditions (as we discussed), we preserve 3d ${\cal N}=2$ superconformal symmetry, so what we would find on the boundary (with a coefficient proportional to $\theta$) is actually the supersymmetric completion of the Chern-Simons term. This completion includes extra quadratic terms involving the background fields coupling to the gauginos, and the background fields coupling to the real and imaginary parts of the scalars in the bulk ${\cal N}=2$ vector multiplet. These extra terms modify the boundary conditions on these fields by an effect which is exactly equivalent to a $U(1)_R$ transformation. Recall that the classical ${\cal N}=2$ theory has a $U(1)_R$ symmetry, under which the gauginos carry charge one and the scalars charge two. This symmetry is anomalous, meaning that a $U(1)_R$ transformation is not a symmetry, but it does leave the physics invariant if we accompany it by an appropriate shift in the $\theta$ angle. On AdS$_4$, this $U(1)_R$ is explicitly broken by the boundary conditions. However, the effect of the supersymmetric Chern-Simons term is that changing the $\theta$ angle is accompanied by a $U(1)_R$ rotation of the fields, in such a way that the bulk physics is invariant. This rotation reproduces the required dependence of the vacuum expectation values of scalar fields in the supersymmetric vacua on the $\theta$ angle.



\end{document}